\documentclass[pre,superscriptaddress,twocolumn,a4paper,showpacs]{revtex4-1}
\usepackage{graphics}
\usepackage{amssymb}
\usepackage{wasysym}
\usepackage{latexsym}
\usepackage{graphicx}
\usepackage{epsfig}
\usepackage{subfigure}

\begin{document}

\title{A stochastic model for the influence of social distancing on loneliness}

\author{Jos\'e F.  Fontanari}
\affiliation{Instituto de F\'{\i}sica de S\~ao Carlos,
  Universidade de S\~ao Paulo,
  Caixa Postal 369, 13560-970 S\~ao Carlos, S\~ao Paulo, Brazil}

\begin{abstract}
The short-term    economic consequences of the critical measures employed to curb the transmission of  Covid-19 are all too familiar, but the consequences of  isolation and loneliness resulting from those measures  on the  mental well-being of the population   and their ensuing long-term economic effects are largely unknown.
  Here we offer a stochastic agent-based model  to investigate social restriction measures in a community  where the  feelings of  loneliness of the agents  dwindle   when they are socializing and grow when they are alone. In addition, the intensity of those feelings, which are measured by a real variable that we term degree of loneliness,  determines whether the agent will seek social contact or not.  We find that
  decrease of  the number,  quality or  duration  of social contacts lead the community to enter a regime of burnout in which the  degree of loneliness diverges, although the number of lonely agents at a given moment   amounts to only a fraction of the  total population. This regime of mental breakdown is separated from the  healthy regime, where the degree of loneliness is finite, by   a  continuous phase transition. We show that the community dynamics is described  extremely well by a simple   mean-field theory so our conclusions can be easily verified for different scenarios and parameter settings.  The appearance  of the burnout regime  illustrates neatly  the side effects of  social distancing, which give to  many of us the choice between physical infection and mental breakdown.

\end{abstract}

\maketitle

%
%-----------------------------------------------------
\section{Introduction} \label{sec:intro}
%-----------------------------------------------------
 %
 
   Even before the Covid-19 pandemic, the World Health Organization declared  social disconnection a major public health challenge, since the lonely and socially isolated  face heightened morbidity and mortality risks: today,  lonely people are 30\% more likely to die early than less lonely ones \cite{Leigh2017,Alberti2019,Courtet2020}. To address this crisis and prompted by reports that about 13\%  of its population feel lonely some or all of the time and that this social  disconnection  may be costing its economy 32 billion pounds a year \cite{Cox2017},  the United Kingdom created a Ministry of Loneliness in 2018.  Japan followed suit in 2021.
   
 Against this current,  the Covid-19 pandemic  has brought unprecedented efforts to enforce social distancing and quarantining all over the world.   While these measures are  unarguably pivotal to preventing the spread of this disease, they will undoubtedly have consequences for mental health  in both the short and long term. For many people today, the choice is between physical infection and mental breakdown \cite{Miller2020,Galea2020,Saltzman2020}.  Understanding those consequences from  a quantitative perspective is of sufficient importance to merit a fraction of the 
 attention spent on the  mathematical and computational  modeling of the  Covid-19 transmission dynamics (see, e.g., \cite{Bellomo_20}).    In fact, given the well-established influence of positive affect on  cognitive function and hence on productivity (see, e.g., \cite{Isen_87,Oswald_15}),  the long-term socio-economic implications of the Covid-19 pandemics may be  far more serious than the prognoses of the economic pundits  \cite{Nicola2020}.
 
 Accordingly, to address the impact of social distancing on individual and population level mental health we  use an   agent-based model to simulate a community dynamics where the feelings of loneliness of an  agent is measured by a real variable - the loneliness degree - that determines   the propensity of the agent  to initiate a  social interaction (or conversation) as well as to terminate an ongoing interaction. The loneliness degree increases when the agent is alone and decreases when it 
 is socializing, in agreement with the  findings that positive affect  increases significantly after social interaction \cite{Phillips_67,McIntyre_91,Cacioppo_09}. Social (or, more correctly, physical) distancing is modeled by controlling  the number of attempts an agent makes to find a conversation partner.  More importantly, our model takes into account the quality of the social interaction that is measured by the rate at which the degree of loneliness decreases during a social interaction.  In fact, a unique characteristic of the current  pandemic is  the wide access to technology that, in principle, might help buffer loneliness and isolation  \cite{Smith2018,Banskota2020}. However, evidence of heightened psychological problems amongst  the youth in the wake of this pandemic \cite{Liang2020} indicates that the abundance of virtual social contacts may have actually little or even negative impact on the feelings of loneliness \cite{Miller2018} as the so-called `Zoom fatigue'  illustrates so nicely. Hence the quality of the social interactions matters, regardless of whether they are virtual or physical  \cite{Moorman2016}.

Our approach builds on an agent-based model proposed to address  the influence of social distancing   on productivity \cite{Peter2021}.   However, in addition to the agent-based simulations, here we offer an analytical mean-field approximation that describes the simulation results very well and allows our results and conclusions to  be easily  verified for distinct  parameter settings.  Our main finding is that decrease of  the number,  quality or  duration  of social contacts lead the community to enter a regime of burnout in which the  degree of loneliness diverges.
This regime of mental breakdown is separated from the  healthy regime, where the degree of loneliness is finite, by   a  continuous phase transition in the sense that the proportion of lonely agents in the community changes continuously when transitioning between those regimes.   This unexpected  threshold phenomenon highlights 
our unfamiliarity with the  mental health  consequences of isolation and loneliness resulting from the social distancing measures.

%
%-----------------------------------------------------
\section{Model}\label{sec:model} 
%-----------------------------------------------------
%

We consider a community composed of $N$ agents that can either interact socially or  remain alone depending on their feelings of loneliness.  The feeling of loneliness of an agent, say agent   $k$,  is measured by its loneliness degree $L_k \in \mathbb{R}$ that, in turn, determines   the propensity  of this agent to seek and engage in social interaction as well as to end an ongoing interaction. Here we assume that lonely people feel the need for company \cite{Alberti2019}. In addition, we  assume that $L_k$  is affected differently depending on whether agent $k$  is alone or interacting with another member of the community. This assumption introduces a feedback  between loneliness and behavior that is responsible for the nontrivial results  of the model dynamics.

If agent $k$ is alone then the probability that it will attempt to instigate a conversation with another lonely agent is given by $p_k = p(L_k)$, where $p(x) \in [0,1]$ is an arbitrary function. When the lone  agent $k$ decides to instigate a conversation, it selects  a number $m$ of  contact attempts, where $m =0, 1, \ldots$ is a random variable drawn from a Poisson distribution of parameter $q$.  In each contact attempt, a  mate is selected at random among the $N-1$ agents in the  community and, in case the selected agent is alone at that moment,  a conversation  is initiated   and the  agent $k$ halts its search for a mate.  If none of the $m$ selected agents are alone, then the  attempt of the agent to socialize fails and it remains alone. 
%In the context of the 2020 coronavirus pandemic, $q$ is the leading parameter of the model as    the  mean number of  attempts to interact  socially  was  considerably curtailed  by the enforced social distancing measures. 
A conversation or social interaction involves  two agents only and  the agent that is approached by agent $k$  is obliged to accept the interaction, regardless of  its loneliness degree. This pro-social behavior is chosen in order to not further complicate the model, but it can be justified in terms of social norms especially during the  current pandemic when there is a pressure to talk to everyone because one  worries that they are lonely and one does not want to turn them down. Of course,  
this pro-social behavior is one of the causes of the Zoom fatigue. If agent $k$  is socializing then the probability that it will unilaterally interrupt the conversation is given by 
$r_k = r(L_k)$, where $r(x)\in [0,1]$   is another arbitrary function. In addition,  the rate of change of the loneliness degree of agent $k$  is determined by the function $M_a(L_k)  \in \mathbb{R} $ if it is alone and by  the function  $ M_s(L_k)  \in \mathbb{R} $ if it is socializing.

The asynchronous evolution of the community of $N$ agents at time $t$  proceeds as follows. In the  time interval $\delta t$, 
we pick an agent at random, say agent $k$, 
and  check if it is  alone or socializing.  In case it is alone, we change its loneliness degree according to the prescription
\begin{equation}\label{Ma_1}
L_k^{t+\delta t} = L_k^{t} + M_a (L_k) \delta t
\end{equation}
and test if it  will attempt to  initiate 
a conversation using the   socializing probability $p_k = p(L_k^t)$.  As mentioned before, this attempt involves the selection with replacement of at most $m$  members of the   community until another lone agent is found. In case agent $k$ is socializing, we change its loneliness degree according to the prescription
\begin{equation}\label{Ms_1}
L_k^{t+\delta t} = L_k^{t} + M_s (L_k) \delta t
\end{equation}
 and then check if it will  terminate the conversation
 using the termination probability  $r_k = r(L_k^t)$. In case it does, both agent $k$ and its mate  become lonely at time $ t + \delta t$.
  As usual in such asynchronous update scheme, we choose the time increment as $\delta t = 1/N$ so that during the increment from $t$ to $t+1$ exactly  $N$, though  not necessarily  distinct,  agents are chosen to follow the  update rules.  
  
 To avoid misinterpretations  of the behavioral rules described  above, it is convenient to write them in a more formal manner. For  instance, given that agent $k$ is alone at time $t$, the probability that it will remain alone at time $t+\delta t$ is
\begin{eqnarray}\label{a_a_1}
Q_k(a,t+\delta t \mid  a, t) & = &  \frac{1}{N} \left [ 1-p_k + p_k  e^{ -q  (N_a^t-1)/(N-1)}    \right ]  \nonumber \\
&  & \mbox{} 
 + \frac{1}{N}   \sum_{i \in \mathcal{L}_a^t; i \neq k}   [ 1-p_i    + p_i e^{-  q/(N-1)}    ]    \nonumber \\
 &  & \mbox{}  + \frac{N-N_a^t}{N},
\end{eqnarray}
where $N_a^t$  and $ N - N_a^t$ are the numbers of lone and socializing agents  at time $t$, respectively.  The sum in the second term of the rhs of this equation is over the subgroup of lone  agents $\mathcal{L}_a^t$, except  agent $k$,   at time $t$. For notational simplicity, we have omitted the time dependence of $p_k$. The first term of   the rhs of equation  (\ref{a_a_1})  accounts for the possibility that agent $k$ is the agent selected for update, which is an event that  happens with probability $1/N$. In this case there are two possibilities: agent $k$ decides to remain alone, which happens with probability $1-p_k$ or decides to instigate a conversation but fails to find another lone agent, which happens with probability 
\begin{equation}
p_k \sum_{m=0}^\infty     e^{-q} \frac{q^m}{m!} \left ( 1 -  \frac{N_a^t-1}{N-1} \right )^m  = p_k  e^{-q (N_a^t-1)/(N-1)}  .
\end{equation}
The second term of  the rhs of equation (\ref{a_a_1}) accounts for the possibility that a lone agent $i \neq k$  is chosen for update  and that this agent either decides to remain alone, which has probability $1-p_i$, or instigate a conversation  with any other agent but agent $k$, which has probability 
\begin{equation}
p_i \sum_{m=0}^\infty     e^{-q} \frac{q^m}{m!} \left ( 1 -  \frac{1}{N-1} \right )^m  = p_i  e^{- q/(N-1)}  .
\end{equation}
Finally, the third term of  the rhs of  equation (\ref{a_a_1}) accounts for the possibility that the agent selected for update in the time interval $\delta t$ is
one of the $N-N_a^t$  agents that are socializing at time $t$. Since a lone agent at time $t$ can either remain alone or start socializing at time $t+ \delta t$, the probability that the lone agent $k$ at time $t$  starts socializing during the time interval $\delta t$ is readily obtained from the complement rule of probability,
\begin{eqnarray}\label{s_a_1}
 Q_k (s,t+\delta t \mid  a, t)  & = &  \frac{p_k}{N}  \left [ 1-   e^{-q  (N_a^t-1)/(N-1)}   \right ]
\nonumber \\
 &  & \mbox{} 
 +   \sum_{i \in \mathcal{L}_a^t; i \neq k}  \frac{p_i}{N}   \left [  1-  e^{ - q/(N-1)   } \right ] . \nonumber \\
 &  & \mbox{}
\end{eqnarray}

Next, we assume that agent $k$ is socializing  with agent $k'$ at time $t$. The  probability that this interaction continues during the time interval $\delta t$ is simply
\begin{equation}\label{t_t_1}
 Q_{k,k'} (s,t+\delta t \mid  s, t)  =  \frac{1}{N} (1-r_k) +  \frac{1}{N} (1-r_{k'})  + \frac{N-2}{N} ,    
\end{equation}
where we have omitted the time dependence of $r_k$ and $r_{k'}$.
Here the first two  terms of the rhs of this equation account for the  events that agents $k$ and $k'$ are selected for update and they choose not to interrupt their conversation. The last term of the rhs of equation (\ref{t_t_1}) accounts for the event that any other agent, aside from  $k$ and $k'$, is selected for update at time $t$. As before, the event that $k$ and $k'$ will terminate their conversation during the time increment $\delta t$ is complementary to the event that they will continue the conversation, i.e., 
\begin{equation}\label{a_s_1}
Q_{k,k'} (a,t+\delta t \mid  s, t)  = \frac{1}{N} \left ( r_k + r_{k'} \right ) .
\end{equation}

To conclude the set up of our model, two remarks are in order. First, we note that equations (\ref{a_a_1})  and (\ref{t_t_1}) are probabilities of events that occur in the time interval $\delta t $ and so they should be proportional to $\delta t$. This  is in fact the case provided  we set $\delta t = 1/N$. Here we will not consider the unrealistic limit of infinitely large communities  $N \to \infty$ which would correspond  to a continuous-time  model of  the community dynamics. Second, equation (\ref{t_t_1}) introduces a short-time correlation between the loneliness degrees and behaviors of agents  $k$ and $k'$ that hinders an exact analytical approach to solve the model. However,  in the next section we will  set forth a simple mean-field approximation that yields a remarkably good description of some macroscopic features of the community dynamics.

%
%-----------------------------------------------------
\section{Mean-field approximation}\label{sec:MF} 
%-----------------------------------------------------
%

Here we offer a simple but surprisingly effective analytical approximation to the agent-based model described in the previous section. 
A macroscopic quantity of interest is the number   of lone agents $N_a^t$ in the community at time $t$. In  the time interval $\delta t $ this random variable can increase by two agents, decrease by two agents or remain the same. More pointedly, given
 $N_a^t$ and the  loneliness degrees  $L_k^t, k=1, \ldots, N $  at time $t$, the probabilities of those events are
\begin{eqnarray}
P \left ( N_a^{t+\delta t}  = N_a^t + 2   \right)  & = &    \sum_{k \in \mathcal{L}_s^t} \frac{r_k}{N} \\
P \left ( N_a^{t+\delta t}  = N_a^t - 2     \right ) & =&  \sum_{k \in \mathcal{L}_a^t}   \frac{p_k}{N}  \left [ 1 -  e^{-q  (N_a^t-1)/(N-1)}   \right ] \nonumber \\
&  &  \mbox{}
\end{eqnarray}
and $P \left ( N_a^{t+\delta t}  = N_a^t  \right ) = 1 - P \left ( N_a^{t+\delta t}  = N_a^t + 2 \right )  - P \left ( N_a^{t+\delta t}  = N_a^t - 2 \right )$.  
 Hence the expected number of lone agents at time $t+ \delta t$ given that there are $N_a^t$ lone agents at time $t$ is
\begin{eqnarray}\label{ex_Na}
\langle  N_a^{t+\delta t} \rangle  &  = &  N_a^t + 2 P \left ( N_a^{t+\delta t}  = N_a^t + 2 \right )  \nonumber \\
&  & \mbox{} -2  P \left ( N_a^{t+\delta t}  = N_a^t - 2 \right )  .
\end{eqnarray}

In a similar vein, we can write the expected loneliness degree of agent $k$ at $t + \delta t$ as
\begin{eqnarray}
\langle L_k^{t + \delta t} \rangle  &  = &  \left [ L_k^{t} + M_a (L_k^t) \delta t \right ] \frac{1}{N} \frac{N_a^t}{N}  \nonumber \\
& &  \mbox{}  + 
\left [ L_k^{t} + M_s (L_k^t) \delta t \right ] \frac{1}{N} \frac{N-N_a^t}{N} + L_k^t  \frac{N-1}{N} \nonumber \\
 &  = &   L_k^{t}  +  \frac{N_a^t}{N} \left [  M_a (L_k^t) -  M_s (L_k^t) \right ]  \frac{\delta t}{N}  
 \nonumber \\
&   & \mbox{}  + M_s (L_k^t) \frac{\delta t}{N}, 
\end{eqnarray}
where we have used that the probabilities that agent $k$ is alone or socializing at time $t$ are $N_a^t/N$ and $(N-N_a^t)/N$, respectively. 

To proceed further  we make the usual mean-field assumption $N_a^t \approx \langle  N_a^{t} \rangle \equiv N \eta^t $ and $L_k^t \approx \langle  L_k^{t} \rangle  $ (see, e.g., \cite{Huang_63}).  In addition, we assume that the mean loneliness degree is the same for all agents, i.e., $ \langle  L_k^{t} \rangle  = \langle  L^{t} \rangle \equiv l^t$. These assumptions suffice for writing the mean-field version of  the community dynamics,
\begin{eqnarray}\label{mf_1}
\eta^{t+\delta t}  & =  & \eta^t + 2  (1-\eta^t) r(l^t)  \delta t  \nonumber \\ 
&  & \mbox{} - 2 \eta^t p (l^t)   \left [ 1 -  \exp \left (-q  \frac{\eta^t-1/N}{1-1/N}  \right ) \right ]   \delta t 
\end{eqnarray}
\begin{equation}\label{mf_2}
l^{t + \delta t}  =  l^{t} + \left [ \eta^t  \left ( M_a (l^t) - M_s (l^t)   \right ) + M_s (l^t)  \right ] \frac{\delta t}{N} 
\end{equation}
where we have used $\delta t = 1/N $ in equation (\ref{mf_1}) to stress the  incremental nature of the intensive variable $\eta^t$.

In the case equation (\ref{mf_2}) has a fixed point $l^{t+\delta t} =  l^{t} = l^*$, the equilibrium fraction of lone agents   
$\eta^{t+\delta t} =  \eta^{t} = \eta^*_h$ is given by
\begin{equation}\label{eta*}
\eta^*_h = \frac{M_s (l^*)}{M_s (l^*) - M_a (l^*)}
\end{equation}
with $l^*$ given by  the solution of the transcendental equation
\begin{equation}\label{l*}
- \frac{M_a (l^*) r(l^*)}{M_s (l^*) p(l^*)} = 1 -  \exp \left (-q  \frac{\eta^*_h-1/N}{1-1/N}  \right ) .
\end{equation}
The subscript $h$ in our notation for the equilibrium fraction of lone agents $\eta^*_h$  stands for healthy since  $l^*$ is finite for this solution. The condition  $\eta^*_h \in [0,1] $  requires that  either $M_a(l^*) < 0$ and $M_s(l^*) >0$ or $M_a(l^*) > 0$ and $M_s(l^*) <0$.  Since $l^t$ measures the degree of loneliness of a generic agent we will assume that $M_a(l^t) > 0$ and $M_s(l^t) < 0$ which, according to equations (\ref{Ma_1}) and  (\ref{Ms_1}),  means that the loneliness degree of an agent increases when it is alone and decreases when it is socializing.

An  interesting situation occurs when  equation (\ref{l*})  has  no solution so that  $l^t \to  \infty $  in the limit $t \to \infty$. This divergence characterizes a burnout regime  where the equilibrium fraction of lone agents $\eta_b^*$  is given by the solution of the equation
\begin{equation}\label{etab*}
 \lim_{l^t \to \infty}  \frac{r(l^t)}{p(l^t)} =   \frac{\eta^*_b}{1-\eta^*_b}  \left [  1 -  \exp \left (-q  \frac{\eta^*_b -1/N}{1-1/N}  \right ) \right ], 
\end{equation}
which is obtained from equation (\ref{mf_1}) by setting $\eta^{t+\delta t} = \eta^t = \eta^*_b$ and the subscript $b$ in $\eta^*_b$ stands for burnout.

%
%-----------------------------------------------------
\section{Results}\label{sec:res} 
%-----------------------------------------------------
%
In the previous sections, we  have made no assumptions on the probability functions $p(l) $ and $r(l)$ that determine the effect of the loneliness degree $l$ on the behavior of the agents. The functions $M_a(l) > 0$ and $M_s(l) < 0$ that determine the changes on the loneliness  degree of  lone and  socializing agents, respectively, were  left unspecified too.  However, in order to simulate the model we need to specify those functions.  Here we assume that the
propensity to instigate a conversation is a decreasing function of the loneliness degree of the agents,
\begin{equation}
p(l) = \frac{1}{2} \left [ 1 +  \tanh (\beta l ) \right ],
\end{equation}
where $\beta \geq 0$ is a parameter that determines the influence of the loneliness  on the behavior of the agent. For instance, for $\beta =0$, the loneliness has no effect on an agent's decision to instigate or not a conversation, whereas for $\beta \to \infty$ a lone agent will always attempt to socialize when $l > 0$.  Moreover, we assume  that the probability that a socializing agent terminates a conversation does not depend on its  loneliness degree, i.e., $r(l) = r \in [0,1]$, since there are many external factors that may result in the interruption of a conversation, in contrast to the longing to socialize, which is most likely fed by internal factors \cite{Alberti2019}. Finally, for the sake  of simplicity, we assume that the rates of change of the loneliness degrees are constant, i.e.,  $M_a (l) = a > 0$ and $M_s (l) = - s < 0$.  Without loss of generality,  we set $a=1$, since this parameter can be removed from our equations by a proper rescaling of $L_k$, $s$ and $\beta$.

With the above choices we can rewrite equations (\ref{eta*}) and (\ref{l*}) and obtain explicit expressions for $\eta^*_h $ and $l^*$, viz., 
\begin{equation}\label{eta**}
\eta^*_h = \frac{s}{1+s}
\end{equation}
\begin{equation}\label{l**}
l^* = \frac{1}{2 \beta} \ln \left ( \frac{\Lambda}{1-\Lambda} \right) 
\end{equation}
where
\begin{equation}\label{lambda}
 \Lambda =  \frac{r/s}{ 1 -  \exp \left (-q  \frac{s/(1+s)-1/N}{1-1/N}  \right ) }.
\end{equation}
This fixed point exists provided that $\Lambda < 1$ and a necessary (but not sufficient) condition for this happening is $r/s < 1$. In fact,  a small value of $r$ implies that the conversations last longer and a large value of $s$ implies that they  bring about a substantial  diminution of the feelings of loneliness.  (We recall that the comparison baseline  of $s$ is the increment of the loneliness degree of the lone agents, viz.,  $a=1$.) Hence,   the lesser the rate $r/s$, the healthier the agents,  provided, of course, that they can find conversation partners whenever they need one.

\begin{figure}[t] 
\centering
 \includegraphics[width=.8\columnwidth]{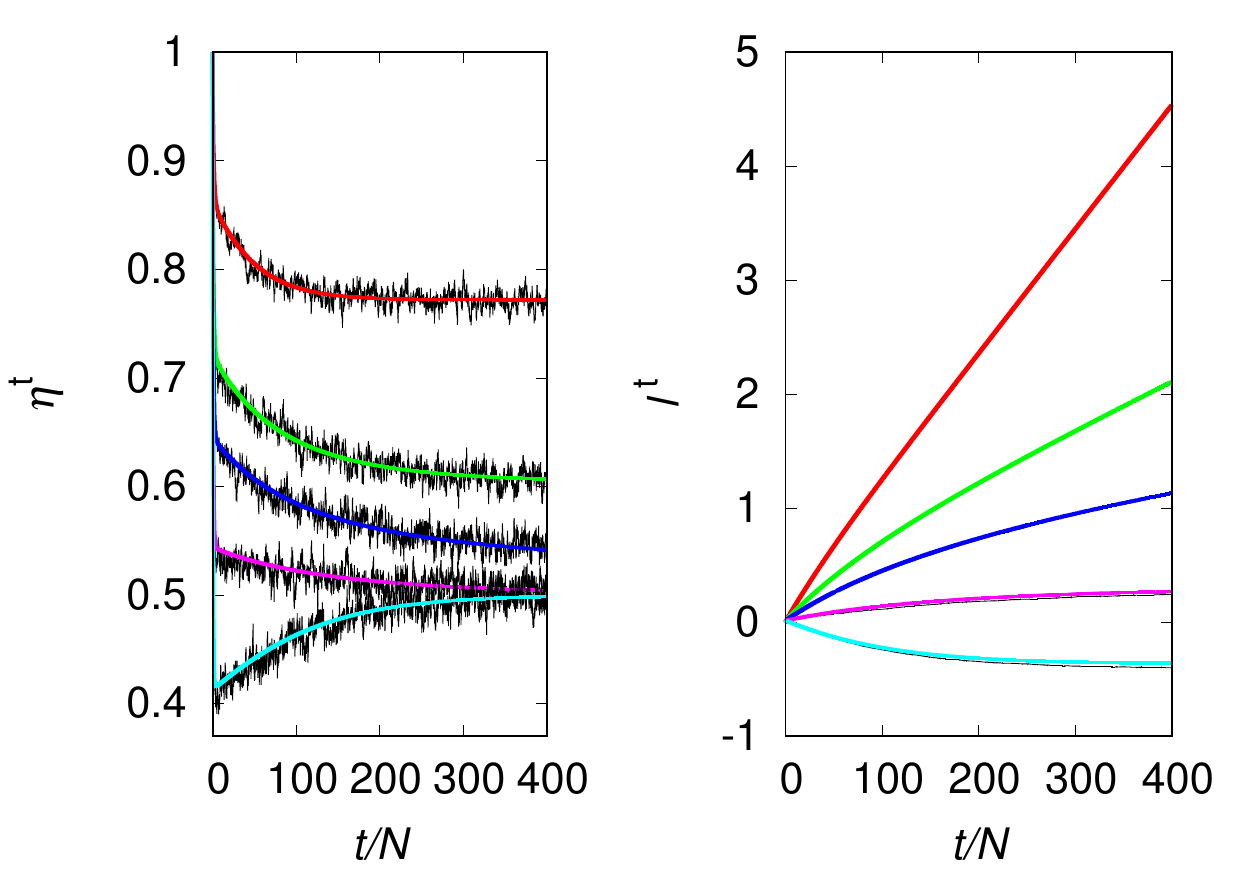}  
\caption{Time evolution of the mean fraction of lone agents $\eta^t$  (left panel) and mean loneliness per agent  $l^t$ (right panel) for a population of  size $N=50 $ and  mean number of contact attempts (from top to bottom) $q=0.1, 0.3, 0.5, 1$ and $3$.  The other parameters are $r=0.25$, $s=1$ and $\beta = 1$. The critical point occurs at $q_c = 0.587$.  The colored thick lines are the mean-field predictions and the black thin lines are  the averages over $10^2$  independent agent-based simulations. The initial conditions are $N_a^0=N$ and $L_k^0 = 0, k=1, \ldots, N$ so that $\eta^0 = 1$ and 
$l^0 = 0$.
 }  
\label{fig:1}  
\end{figure}

What happens in the case that $\Lambda \geq 1$?  Iterating equations (\ref{mf_1}) and (\ref{mf_2})  with $\delta t = 1/N$ (see figure \ref{fig:1})  we find that $l^t \to  \infty $  in the limit $t \to \infty$ whereas $\eta^t$ tends to the finite value $\eta^*_b$  given by equation (\ref{etab*}), which reduces to
\begin{equation}\label{eta2}
r = \frac{\eta^*_b}{1-\eta^*_b}  \left [  1 -  \exp \left (-q  \frac{\eta^*_b -1/N}{1-1/N}  \right ) \right ]
\end{equation}
since  $\lim_{t \to \infty} p(l^t) = 1$ and $r(l^t) = r$.  We note that for $\Lambda = 1$ equation (\ref{eta2}) reduces to equation (\ref{eta**}), i.e., $\eta^*_b = \eta^*_h$,  so that the transition between the healthy and burnout regimes is continuous regarding  the asymptotic mean fraction of lone agents. In fact, the  condition $\Lambda = 1$ determines the critical  value of the mean number of attempts to make a social contact 
\begin{equation}\label{qc}
q_c = - \frac{1-1/N}{s/(1+s)-1/N} \ln \left ( 1 - r/s \right )
\end{equation}
with $r/s < 1$. The healthy regime occurs for $q > q_c$ (i.e., $ \Lambda < 1$)  and the burnout regime for $q \leq q_c$  (i.e., $ \Lambda \geq 1$).   In the case that $r/s > 1$, the model exhibits  the  burnout regime only with  $\eta^*_b$ given by equation (\ref{eta2}). In this case,  the equilibrium fraction of lone agents  does not depend on $s$.

%-----------------------------------------------------
%-----------------------------------------------------

\begin{figure}[t] 
\centering
 \includegraphics[width=.8\columnwidth]{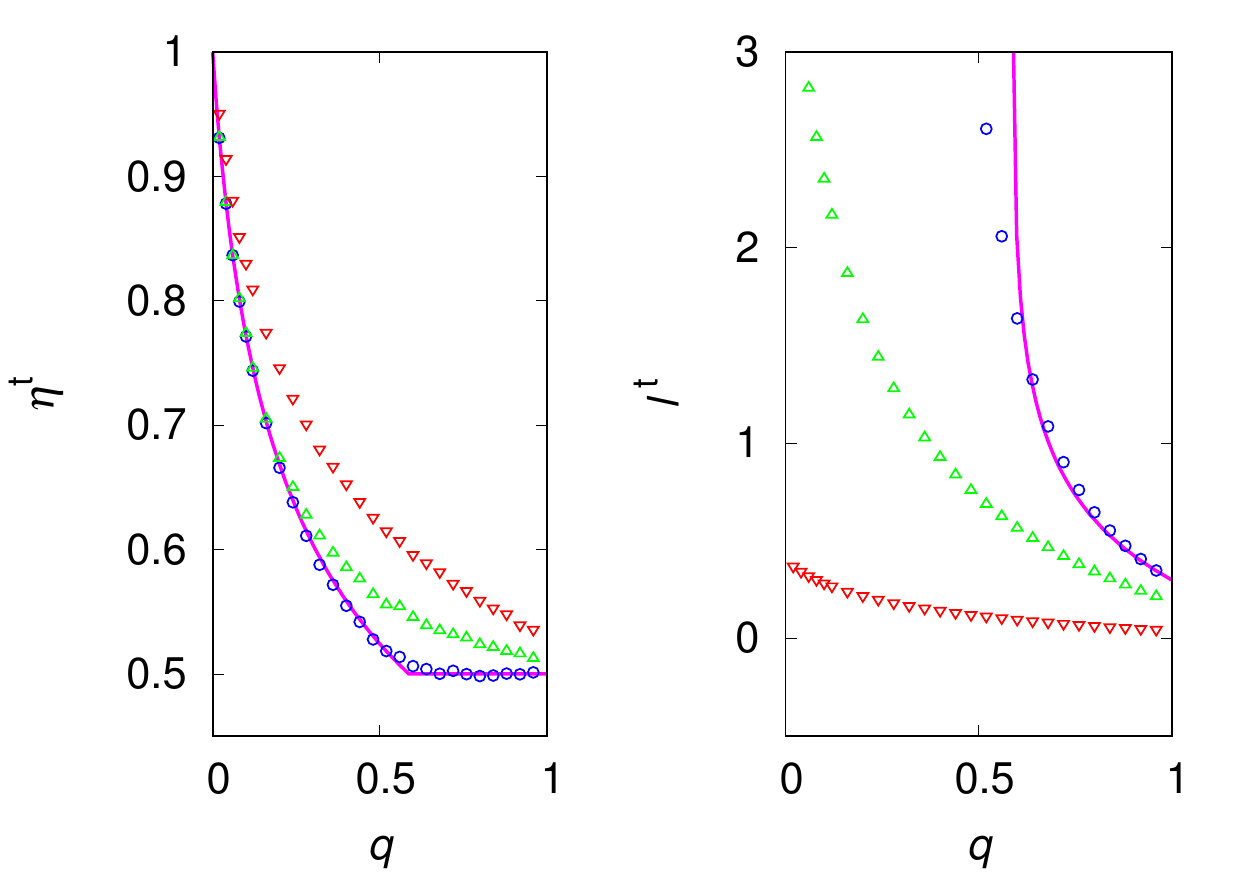}  
\caption{ Mean fraction of lone agents $\eta^t$  (left panel) and mean loneliness per agent  $l^t$ (right panel) evaluated at $t=10^3$ ($\triangledown$), $t=10^4$ ($\triangle$) and $t=10^5$ ($\circ$) as functions of the mean number of contact attempts $q$. The symbols represent the averages  over $10^4$ independent agent-based simulations. The solid  lines are the mean-field predictions for the limit $t \to \infty$. The critical point occurs at $q_c = 0.587$.  
The other parameters are $N=50 $, $r=0.25$, $s=1$ and $\beta = 1$.    
 }  
\label{fig:2}  
\end{figure}

In figure \ref{fig:1},  we show the time evolution of $\eta^t$ and $l^t$  for the simulation of the agent-based model as well as for the mean-field approximation. The agreement between them is so remarkable that we have averaged those quantities over  only 100 independent simulations in order to make the differences  noticeable, though with no success in the case of the mean loneliness degree $l^t$. This agreement seems rather puzzling at first sight because the mean-field approximation exhibits a phase transition between the healthy and burnout regimes  that cannot  be observed in the  `finite'  agent-based system of our simulations. In fact, the signatures of the phase transition, viz.,  the discontinuity of the derivative of the asymptotic  value of $\eta^t$ with respect to $q$ and the divergence of the asymptotic value of $l^t$ at  $q=q_c$,   appear  in  the `thermodynamic' limit only.  As just hinted,   the thermodynamic limit in our model is the time asymptotic limit $t \to \infty$ and  since we cannot run infinitely long simulations  we will never see those signatures in our simulation results. In figure \ref{fig:2},  we illustrate this point  by showing  $\eta^t$ and $l^t$ evaluated at times $t=10^3, 10^4$ and $10^5$.  These results indicate that the mean-field fixed points describe  very accurately  the asymptotic  time behavior of the agent-based model.

 %-----------------------------------------------------
%-----------------------------------------------------
 
\begin{figure}[t] 
\centering
 \includegraphics[width=.8\columnwidth]{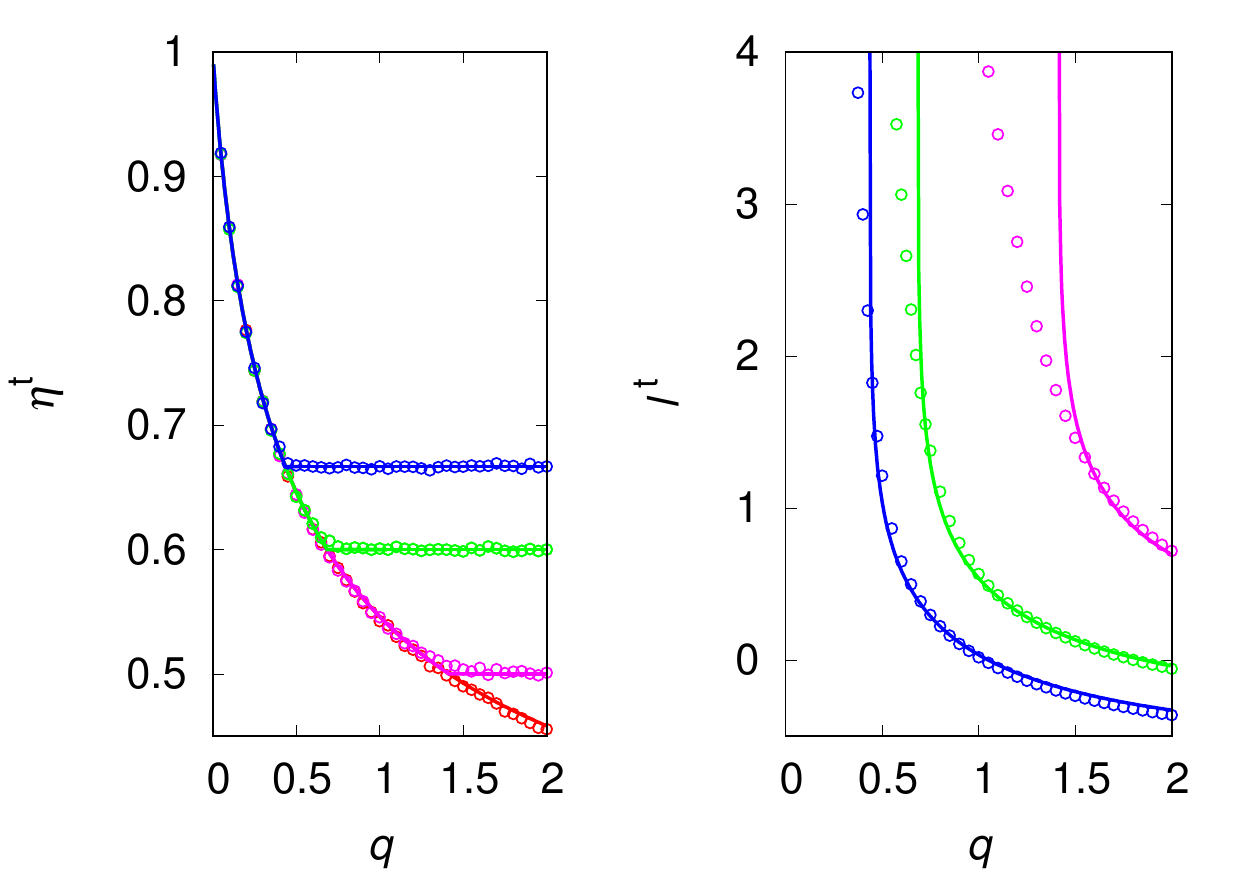}  
\caption{Mean fraction of lone agents $\eta^t$  (left panel) and asymptotic mean loneliness per agent  $l^t$ (right panel)  evaluated at $t=10^5$ as functions of the mean number of contact attempts $q$ for  (left panel from top to bottom) $s = 2,1.5, 1 $ and $0.5$. The symbols represent the averages  over $10^4$ independent agent-based simulations and the solid  lines are the mean-field predictions for the limit $t \to \infty$.
The other parameters are $N=50 $,  $r=0.5$ and $\beta = 1$.  The data for $s=0.5$ is not shown in the right panel  because $l^*$ diverges  in the mean-field approximation and the simulations yield results that are well above the range of the y-axis. 
 }  
\label{fig:3}  
\end{figure}

 In figure \ref{fig:3},  we show that the excellent agreement between the simulation and the mean-field results  holds for other values of the model parameters too.  As pointed out before,  the discrepancies observed near the critical region are most  likely due  to the fact that we evaluate the time-asymptotic quantities  at the finite time $t=10^5$.  In particular, this figure highlights the curious finding that the rate of decrement of the loneliness degree due  socialization $s$ has no influence on the number of lone agents in the burnout regime. The limit $q \to \infty$ guarantees that a lone agent will always find a conversation partner if there is one available. In this case,  the mean-field approximation  yields $\eta_h^* =  s/(1+s)$ and $l^* =(1/2\beta)\ln \left [ (r/s)/(1-r/s) \right ]$  if $r/s < 1$, and $\eta_b^* =  r/(1+r)$  and $l^* \to \infty$ if $r/s \geq 1$.
 
Since the mean-field approximation describes the simulation results   so well,  it is instructive to look into its predictions near the critical point $q_c$ for $r/s < 1$.  In the healthy regime ($q > q_c$) we find
 \begin{equation}
 l^* \approx \frac{1}{2 \beta} \ln ( q- q_c)
 \end{equation}
  and $ \eta^*_h  = s/(1+s) $, whereas in the  burnout regime  ($q < q_c$) we  find
 \begin{equation}
 \eta^*_b \approx \frac{s}{1+s} + \mathcal{A} (1 - \frac{q}{q_c}),
 \end{equation}
 where 
 \begin{equation}
 \mathcal{A}= -  \ln (1-r/s) \frac{s(s-r)(1-1/N)}{q_c s(s-r) + r(1+s)^2(1-1/N)} > 0 .
 \end{equation}
 Hence, if we define the order parameter of the phase transition as $ \rho = \eta_b^* - \eta_h^*$  then $\rho \sim  ( q_c -  q )$ as we approach the  critical point from the burnout regime.
 
%-----------------------------------------------------
%-----------------------------------------------------

At this stage, it is convenient to consider a more microscopic perspective of the community dynamics.  We begin by  pointing out  that, since the $N$ agents are identical regarding the behavioral rules,  the mean proportion of time that, say, agent $k$  spends alone equals the mean fraction of lone agents in the population  for large $t$.   Our simulations indicate  that this equality holds  true only when those quantities are averaged over many independent simulations, hence the adjective `mean' in the above statement. 

%-----------------------------------------------------
\begin{figure}[t] 
\centering
 \includegraphics[width=.8\columnwidth]{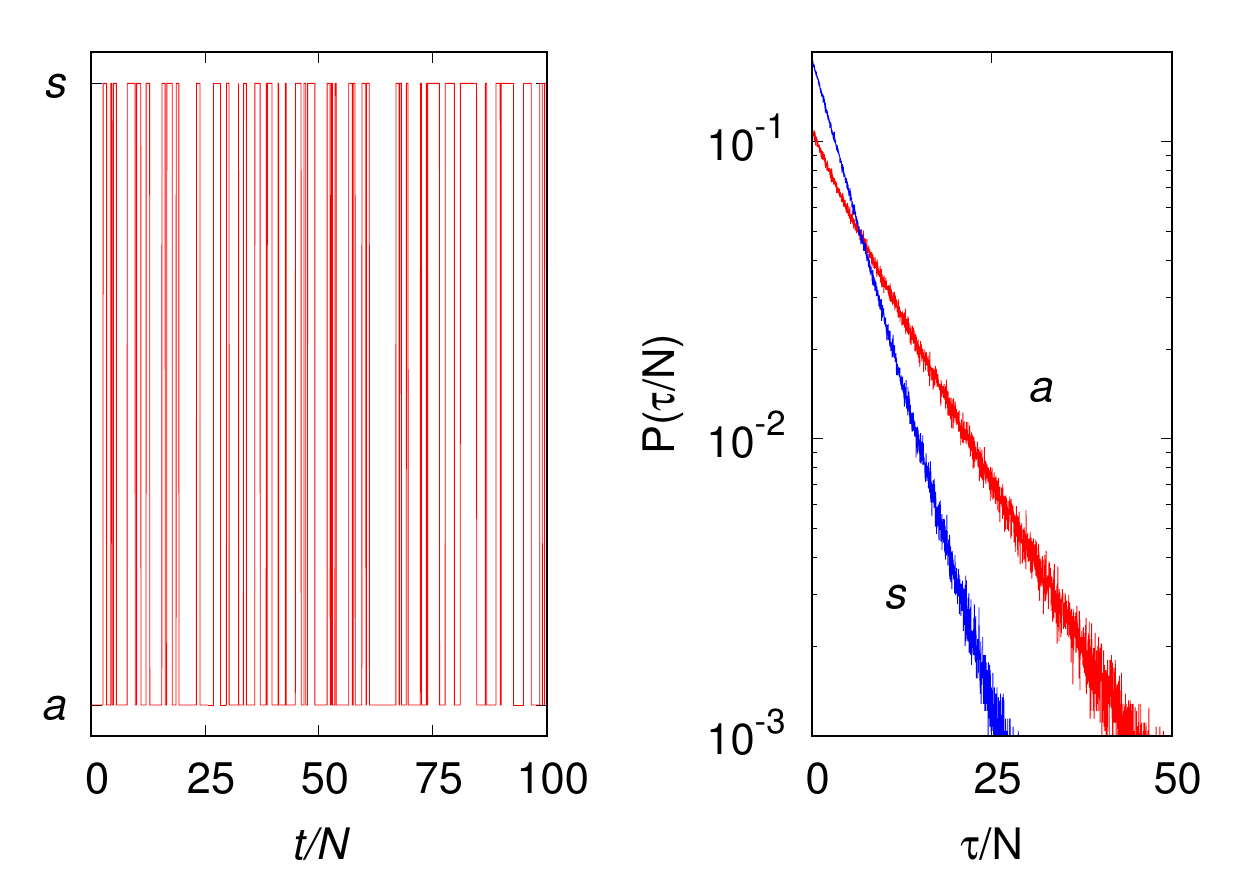}  
\caption{Sequence of flips between the conditions alone (\textsl{a}) and socializing (\textsl{s}) for a single agent in  a single run  (left panel) and  probability distributions of the lengths of time $\tau $ that the  agent spends in states  \textsl{a} and \textsl{s} as indicated   (right panel). 
The  parameters are $N=50 $,  $r=0.1$, $s=2$  and $\beta = 1$.  The exponential probability distributions with means $\langle \tau_\textsl{a} \rangle/N  = 10$  and $\langle \tau_\textsl{s} \rangle/N  = 5$ were obtained  using  $10^4$ independent runs. 
 }  
\label{fig:4}  
\end{figure}
%----------------------------------------------------- 

The left panel of figure \ref{fig:4} shows the flips between  the alone ($\textsl{a}$) and the socializing ($\textsl{s}$) states   experienced by a particular agent during a single run.  The quantities of interest here are the lengths of the periods the agent spends alone $\tau_\textsl{a}$ and socializing $\tau_\textsl{s}$, whose probability distributions are  shown in the right panel of the figure. Since those distributions are observed to be exponential distributions for large $t$, knowledge  of the means  $\langle \tau_\textsl{a} \rangle $ and $\langle \tau_\textsl{s} \rangle $ suffice to describe the random quantities $\tau_\textsl{a}$ and $\tau_\textsl{s}$ in the time-asymptotic limit.  The probability distribution of  $\tau_\textsl{s}$ is  clearly exponential  since  once a couple of agents start socializing the duration of their conversation  does not depend on their previous histories: the conversation is interrupted  when either of the two socializing agents chooses to terminate it, which happens with probability $2r/N$  [see equation (\ref{a_s_1})] so that  
$\langle \tau_\textsl{s} \rangle/N = 1/2r$ \cite{Feller_68}.  As expected, the simulation results perfectly  agree with this prediction (data not shown) which,  we emphasize, does not involve any approximation.  

\begin{figure}[t] 
\centering
 \includegraphics[width=.8\columnwidth]{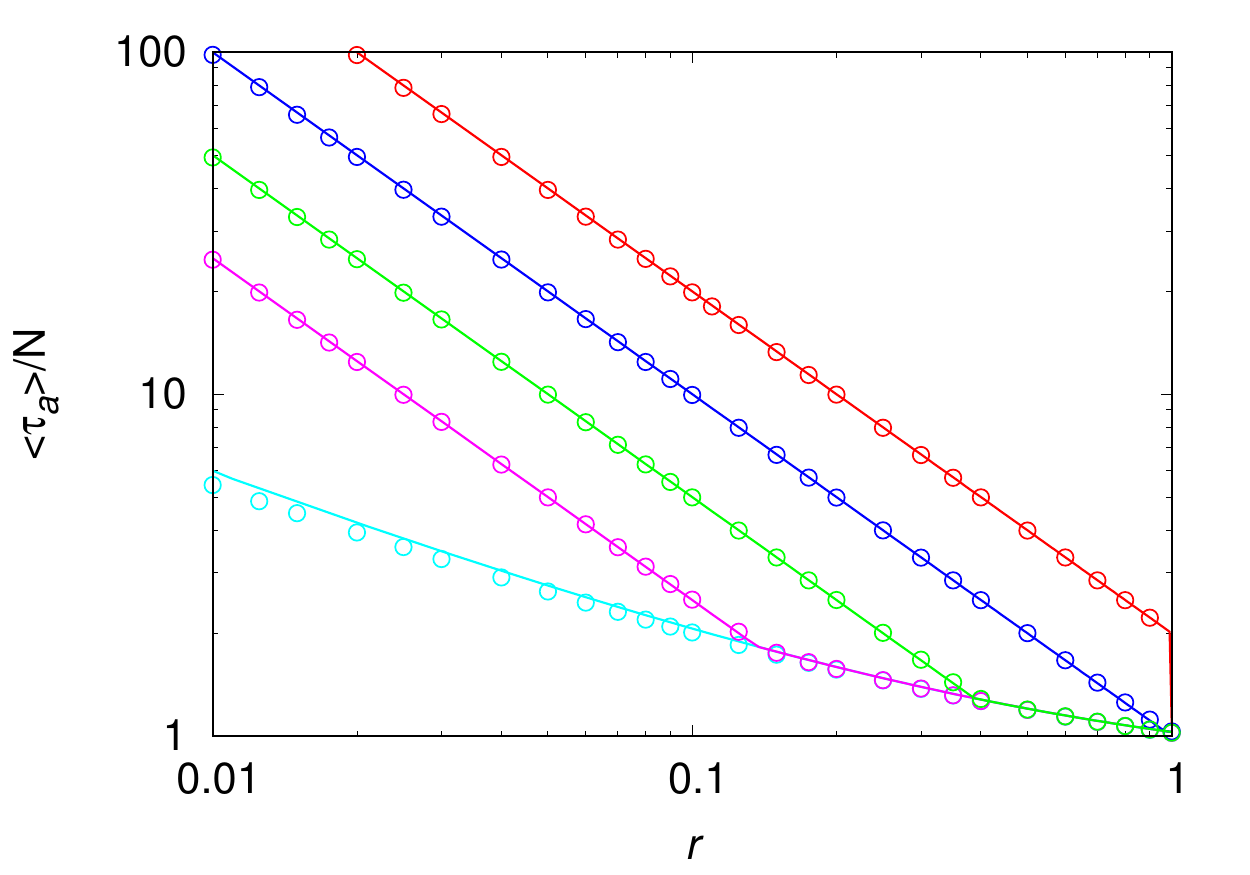}  
\caption{Mean time per agent  that  an agent  spends alone  $\langle \tau_\textsl{a} \rangle/N $  as function of the conversation termination probability $r$ for (top to bottom) $s=4,2,1,0.5$ and $0$.   
  The symbols represent the averages  over $10^3$ independent simulations with the waiting times $\tau_\textsl{a}$   recorded for  $t \in [10^5,10^6]$. The solid  lines are the  predictions of the ansatz (\ref{ansatz}). The  other parameters are $N=50 $,  $q=1$ and $\beta = 1$.
 }  
\label{fig:5}  
\end{figure}

However, the waiting time  $\tau_\textsl{a}$ for a particular lone agent to  start a social interaction  does depend on its previous experiences since the propensity to socialize depends on  its loneliness degree which, in some sense, encapsulates the life  history  of the agent. For instance, if the agent has just terminated a long conversation it  is likely to spend a long time alone before being tempted to socialize again. Nevertheless, our simulations indicate that  the probability distribution of  $\tau_\textsl{a}$ can be described exceedingly well by an exponential distribution. In figure \ref{fig:5} we show $\langle \tau_\textsl{a} \rangle/N$ as function of the conversation termination probability $r$ for fixed $q$.  In this setting,  the phase transition occurs at
\begin{equation}\label{rc}
r_c /s=   1 -  \exp \left (-q  \frac{s/(1+s)-1/N}{1-1/N}  \right ), 
\end{equation}
which corresponds to the condition $\Lambda = 1$ in equation (\ref{lambda}). Since $r_c \leq 1$ there is a value of $s$ above which there is no phase  transition and  the model exhibits  the   healthy regime only. For $q=1$, this happens for $s  > 2.06$.   In contrast to $\langle \tau_\textsl{s} \rangle/N$, the different  time-asymptotic regimes strongly impact  the dependence of $\langle \tau_\textsl{a} \rangle/N$ on $r$,  as  seen in figure \ref{fig:5}. This is expected  because  the probability of finding a conversation partner (and hence of  ending the loneliness period)  depends on the fraction of lone agents $\eta^t$ in the community, which, in turn,  
exhibits rather distinct functional forms in the healthy and burnout regimes,  as illustrated in figure \ref{fig:3}.  

%-----------------------------------------------------
%----------------------------------------------------- 

 We observed that our simulation  results for $\langle \tau_\textsl{a} \rangle/N$ can be  described by a rather simple analytical expression (solid lines in figure \ref{fig:5}) for which we have no explanation.  The probability of the joint event that the lone agent $k$ is chosen for update at time $t$,  decides to instigate a conversation and   succeeds in finding another lone  agent to interact with is
 \begin{equation}
 Q'_{k} =  \frac{ p_k}{N} \left [1-  \exp \left (-q  \frac{\eta^t-1/N}{1-1/N}  \right )  \right ] ,
 \end{equation} 
which is the first term of the rhs  of  equation (\ref{s_a_1}). In the limit of large $t$,  we can replace $\eta^t$  by  its mean-field estimate, namely, $\lim_{t \to \infty} \eta^t = \eta_h^*$ if $r \leq r_c$ and  $\lim_{t \to \infty} \eta^t = \eta_b^*$ if $r > r_c$.  We find that the ansatz 
\begin{equation}\label{ansatz}
\langle \tau_\textsl{a} \rangle/N = 1/(2 N Q'_k) 
 \end{equation} 
offers a perfect  fit for  the simulation results, as shown in figure \ref{fig:5}. In particular, using equation (\ref{l*}) for $r \leq r_c$ we obtain $N Q'_k = r/s$  for large $t$ so that 
$\langle \tau_\textsl{a} \rangle/N = s/2r$. For $r >  r_c$  we obtain $\langle \tau_\textsl{a} \rangle/N  = \eta_b^*/[2r(1-\eta_b^*)] $ where $\eta_b^*$ is the solution of equation (\ref{eta2}). We note that  the  natural guess  $\langle \tau_\textsl{a} \rangle/N = 1/( N Q_k)$  with $Q_k$  given by equation (\ref{s_a_1}) yields qualitatively similar results but   significantly underestimates  the simulation results.

It is  interesting that   both waiting times decrease with increasing $r$. While this result  is obvious for $\tau_s$, it is less apparent  for $\tau_a$. In fact, it is the high  availability of lone agents resulting from short conversations that produces the decrease of $\tau_a$. The reverse is also true:  long socialization periods lead to long periods of loneliness because of the shortage of available partners. In addition, in the healthy regime, the lengths of the loneliness periods increase  with the efficacy of social interactions in reducing loneliness, which is measured by the parameters  $s$. This is expected, since the lesser the degree of loneliness of an agent, the less the probability  that it will seek  social contact.  In the burnout regime,  however, $\langle \tau_\textsl{a} \rangle/N$  does not depend on $s$ provided, of course, that $s$ does not become sufficiently large to allow the transition to the healthy regime.

%
%-----------------------------------------------------
\section{Conclusion}\label{sec:conc} 
%-----------------------------------------------------
%

Since the main measure to curb the spread of SARS-CoV-2 is physical distancing, rather than social distancing, one may argue that internet-based and social media usage may mitigate the feelings of loneliness during the Covid-19 pandemic \cite{Smith2018,Banskota2020}. It is unclear, however, if  use of technology to socialize remotely    can significantly minimize those feelings \cite{Miller2018}. The key issue here is, of course,  the quality of the social interactions. Our model takes this point into account through the parameter $s>0$ that measures the efficacy of the  social interactions  in decreasing feelings of  loneliness. In fact, even if the number of contact attempts is unlimited (i.e., $q \to \infty$) and the community size is very large  (i.e., $N \to \infty$), which is likely the case of social media, an agent  can experience  burnout  in the case that  $s < r $, where $r$ is the probability that the agent ends the  social interaction. We recall that $s < a = 1$ means that the rate  of decrease of the feelings of loneliness when the agent is  socializing  is less than the rate  of increase of those feelings when the agent is alone. It is clear then that $s$ can be used as a proxy for the quality of the social interactions.   Therefore,  our model  describes the effects of  the number of  social contacts  as well as  of the quality of those contacts on loneliness.  Both factors  have been strongly affected  by  the physical distancing and quarantining measures widely implemented to prevent  the spread of Covid-19.  

We find that decrease of  the number,  quality or  duration  of social contacts lead the community to enter a regime of burnout in which the feelings of loneliness  of the agents, measured by the variable $l^t$,  diverge.  This happens through  a  continuous phase transition that separates the healthy from the burnout regimes and that can be identified by the discontinuity of the derivative of the asymptotic  fraction of lone agents  with respect to the parameters of the  model. Since the mean-field approximation reproduces the simulation results very well,   equations (\ref{eta*}),  (\ref{l*}) and (\ref{etab*}) offer  a general formulation of  the  community dynamics where no  assumptions are made on the influence  of   loneliness on  the behavior of the agents, which is determined by   the probabilities $p(l^t)$ and $r(l^t)$, as well as on the effect of that  behavior on the feeling of loneliness, which is determined by the rates $M_a(l^t)$ and $M_s(l^t)$.  In  that sense,  the community dynamics will exhibit a  burnout regime  provided that $\lim_{l^t \to \infty} r(l^t)/p(l^t)$  is nonzero.  
The appearance  of this regime in our model  illustrates neatly  the  side effects of the measures employed to  curb the transmission  of Covid-19 on the population  mental health.

\bigskip

\acknowledgments
I thank Peter Hardy (University of Southampton) for sparking my interest on the modeling of the communal effects of social distancing.
This research  was  supported in part 
 by Grant No.\  2020/03041-3, Fun\-da\-\c{c}\~ao de Amparo \`a Pesquisa do Estado de S\~ao Paulo 
(FAPESP) and  by Grant No.\ 305058/2017-7, Conselho Nacional de Desenvolvimento 
Cient\'{\i}\-fi\-co e Tecnol\'ogico (CNPq).

\end{document}